\documentclass[12pt]{article}

\usepackage{amsmath}
\usepackage{amssymb}
\usepackage{amsfonts}
\usepackage{graphicx}
\usepackage{fullpage}
\usepackage{lineno}
\usepackage[utf8]{inputenc}
\usepackage{mathtools}
\usepackage{fancyvrb}
\usepackage{enumitem}
\usepackage{xcolor}
\usepackage{hyperref}

\newcommand{\modelU}{m_u}
\newcommand{\modelD}{m_d}

\newcommand{\alane}{l} 
\newcommand{\alink}{\ell} 
\newcommand{\alg}{g} 

\newcommand{\lanesupr}{\alane^{up}_r}
\newcommand{\lanesuprp}{\alane^{up}_{r'}}
\newcommand{\lanesdnr}{\alane^{dn}_r}

\newcommand{\GG}{\mathcal{G}}
\newcommand{\RR}{\mathcal{R}}
\newcommand{\HH}{\mathcal{H}}

\newcommand{\Dg}{D_g}
\newcommand{\Dgp}{D^+_g}
\newcommand{\Ur}{U_r}
\newcommand{\Dr}{D_r}
\newcommand{\Uh}{U_h}

\newcommand{\blockedg}{\square^g}
\newcommand{\blockedr}{\square^r}
\newcommand{\blockedh}{\square^h}

\newcommand{\demandgr}{d^g_r}
\newcommand{\demandgrs}{d^g_{rs}}
\newcommand{\supplyh}{s^h}

\newcommand{\demandr}{d^r}

\newcommand{\laneproprh}{\lambda^r_h}
\newcommand{\supplyr}{s^r}
\newcommand{\demandproprh}{\alpha^r_h}

\newcommand{\demandh}{d^h}

\newcommand{\reductionh}{\gamma^h}

\newcommand{\reductionr}{\gamma^r}

\newcommand{\reductiong}{\gamma^g}

\newcommand{\deltagr}{\delta^g_r}
\newcommand{\deltar}{\delta^r}
\newcommand{\deltah}{\delta^h}

\newcommand{\vardxvi}{\sigma^v}
\newcommand{\vardxwi}{\sigma^w}

\newcommand{\linkstates}{\mathcal{S}^\alink}

\usepackage{natbib}

\begin{document}

\title{Open Traffic Models - A framework for hybrid simulation of transportation networks}
\author{Gabriel Gomes}
\maketitle

\begin{abstract}
This paper introduces a new approach to hybrid traffic modeling, along with its implementation in software. The software allows modelers to assign traffic models to individual links in a network. Each model implements a series of methods, refered to as the modeling interface. These methods are used by the program to exchange information between adjacent models. Traffic controllers are implemented in a similar manner. The paper outlines the important components of the method: the network description, the description of demands, and the modeling and control interfaces. We include tests demonstrating  the propagation of congestion between pairs of macroscpoic, mesoscopic, and microscopic models. Open Traffic Models is an open source implementation of these concepts, and is available at https://github.com/ggomes/otm-sim.
\end{abstract}

\section{Introduction}
\label{sec:intro}
Simulation tools are an integral part of transportation planning and research. As described by \citet{Lieberman}, the history of traffic simulators stretches back to the 1950s, and has proceeded alongside developments in the theory of traffic. This trend has continued to the present day: about half of the papers in the Transportation Research Board's 2018 issue on Intelligent Transportation Systems involve a simulation model. Similar numbers apply to recent issues of the IEEE Transactions on Intelligent Transportation Systems.

Transportation models can be classified broadly into macroscopic, mesoscopic, and microscopic models. We adopt here the definitions of \citet{Kessels} for these terms. Macropscopic models do not distinguish individual vehicles, but instead view traffic as a coninuum. This approach originates with the work of \citet{Lighthill} and \citet{Richards}, who coupled the ``fundamental diagram'' of \citet{Greenshields} with the law of conservation of vehicles. In contrast, mesoscopic and microscopic models are vehicle-based. Microscopic models compute vehicle trajectories based on car-following rules. Most models in this category use ordinary differential equations to represent the accelerations of a vehicle as a function of the state of its neighbors. The car-following approach is one of the oldest in transportation modeling, dating back to the work of \citet{Chandler}. It is used today in most micrscopic simulation software, including SUMO \citep{Krajzewicz}, Aimsun (\citeyear{aimsun}), and CORSIM (\citeyear{corsim}).
Another sub-category of micrscopic models are those based on cellular automata. Here space is considered as fundamentally discrete, and the neighborhood of a vehicle consists only of its neighboring cells. These models were introduced to transportation by \citet{Nagel}. Their compatibility with image processing algorithms has lead to extremely fast implementations, such as that of \citet{Korcek}. Mesoscopic models also distinguish individual vehicles, however their movements depend on aggregate quantities such as capacity and jam density, in addition to the states of nearby vehicles. Queueing models fall into this category \citep{DTALite}.

It has long been recognized that none of these model types is superior to all others, but instead each has a domain of application (\citet{Bourrel,Burghout}). For example, models that stem from the kinematic wave theory capture the propagation of congestion waves with fewer parameters than microscopic models. They are more parsimonious for the study of congestion-based interventions, such as ramp metering and variable speed limits on freeways. Conversely, the dynamics of arterial roads is strongly influenced by traffic signals, with congestion wave speed being of secondary importance. Vehicles form queues at intersections, and hence queueing models are a good choice for the study of signal control algorithms. 

\begin{figure}[ht!]
    \centering
    \includegraphics[scale=0.5]{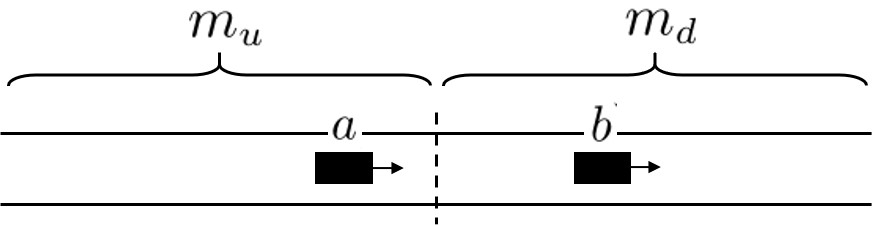}
    \caption{The boundary between two models: $\modelU$ and $\modelD$.}
    \label{fig:ud}
\end{figure}

Observations such as these have motivated the development of \textit{hybrid} approaches, in which different models are applied to different regions of the network. Some early efforts in this area include those of \citet{Bourrel}, \citet{Burghout}, and \citet{Leclercq}. These studies focused primarily on the design of the transition boundary between two \textit{given} models. Figure~\ref{fig:ud} is used for illustration. Here the upstream and downstream regions are managed by models $\modelU$ and $\modelD$ respectively. Vehicles crossing the boundary must be translated from the representation of $\modelU$ into the representation of $\modelD$. If $\modelU$ is microscopic, then the approaching vehicle (vehicle $a$) may require information from a vehicle within $\modelD$ (vehicle $b$), for example to determine its own headway. 

Most authors have approached this problem by creating a ``transition zone'' between $\modelU$ and $\modelD$, where both representations coexist. 
The focus of these studies has not been on the generic case, but rather on how to preserve consistency in the transition zone between two particular models $\modelU$ and $\modelD$. \citet{Burghout} studied the coupling of a mesoscopic model called Mezzo with a microscopic model called MITSIMLab. \citet{Bourrel} proposed the combination of the LWR with a compatible microscopic model, such as \citet{Newell}. This approach is extended by \citet{Leclercq}, who notes that no transitions zone is needed if the models are mutually consistent since, in this case, estimates can be made from upstream information alone. That paper developed a technique for coupling a macroscopic and a microscopic model, both of which were consistent with LWR. 

The goal of the present paper is to develop and demonstrate a more general approach to the problem. First, it is recognized that the function of the transition zone is to preserve information that is needed by one model and which may not be easily obtained from the other. In the example of the figure, if model $\modelD$ is macroscopic, then vehicle $b$ will not have an explicit representation, and thus the headway for vehicle $a$ will be undetermined. A transition zone would guarantee that there is always a vehicle $b$ from which to compute a headway. Here however, we will dispense with the transition zone. Instead it will be the ``responsibility'' of $\modelD$ to provide a reasonable response when queried by $\modelU$ for the position of its upstream-most vehicle. This will not be difficult whenever $\modelD$ is vehicle-based (microscopic or mesoscopic). The case of the cell-transmission model is developed in Section~\ref{sec:fluid} . 

Section~\ref{sec:model_interactions} describes the proposed approach for coordinating the interactions of two models over a boundary. We will see that the methodology captures a wide range of models, including microscopic, mesoscopic, and macroscopic models of first and second order, discretized with a Godunov scheme. The approach allows the arbitrary partitioning of traffic networks into sub-networks, each managed by a different traffic model. This type of architecture, in which link interactions are mediated by a model-agnostic protocol, lends itself well to implementations in distributed memory. A future publication will present the extension to HPC (high performance computing).

The concepts described in this paper have been implemented in the Open Traffic Models project. OTM includes three models representing the three major classes: the cell transmission model (macro), the two-queue model (meso), and Newell's car-following model (micro). Other models can be added, combined, and distributed as plugins, using the OTM interface. OTM also provides several options for controlling (or influencing) traffic. These include fixed actuators such as traffic signals, ramp meters, variable speed limits, as well as on-board devices such as driver information systems and routing apps. The control algorithms that drive these devices can be implemented as plugins as well, using OTM's control algorithm interface.

\section{Preliminaries}
\label{sec:preliminaries}
Two basic requirements for the design of a hybrid traffic simulation interface can be stated: 1) it must \textit{conserve vehicles}, and 2) it must \textit{conserve vehicle characteritics}. The first requirement means simply that vehicles should not be created or destroyed across the interface. The second depends on which characteristics the modeler wishes to track. 
For example, the modeler may assign different performance or emissions characteristics to different populations, such as cars and trucks. Or they may wish to follow different populations of vehicles for the purpose of computing performance metrics. It is also possible that different vehicles have different access to the network infrastructure -- e.g. high-occupancy vehicles can use the HOV lane; connected vehicles and drivers with routing-apps have access to an information service. This is captured with the \textit{vehicle type}. The concept of vehicle type gathers all distinguishing characteristics of the population that we are interested in tracking. Hence the population is split according to vehicle type, and each type is assigned a unique id. 

Vehicles enter the network through a \textit{source}, attached to a link. A source produces a stream of vehicles of the given type (multiple sources can be attached to a single link). The time-varying intensity of the source is specified by the user as a discrete-time profile. Excess demand, that is, demand that exceeds the flow capacity or holding capacity of the link, is held in a limitless buffer. 
The process that produces the vehicles is specified by the user, and also depends on the model that operates on the source link. A vehicle-based model, for example, may create vehicles according to a Poisson process, while a fluid model may use a sequence of independent random variables. 

Each vehicle type is assigned a \textit{routing behavior}. The routing behavior can be either \textit{routed} or \textit{probabilistic}. \textit{Routed} vehicle types are assigned a \textit{route} (a.k.a. a \textit{path}), which is a sequence of links starting with the given source link. These vehicles travel along the path and are removed from the network upon exiting the last link in the sequence. They can be diverted to another path only by a \textit{routing actuator}, such as a routing app. \textit{Probabilistic} vehicles choose their next link at each junction according to turning probabilities, or \textit{split ratios}. These vehicles are removed from the simulation when they exit a terminal link in the network. The split ratios are provided as discrete-time sequences for every junction and vehicle type. They can be modified during the run by an event (e.g. an accident) or a split ratio actuator.

\subsection{Road segments}

\begin{figure}[ht]
    \centering
    \includegraphics[width=6.5in]{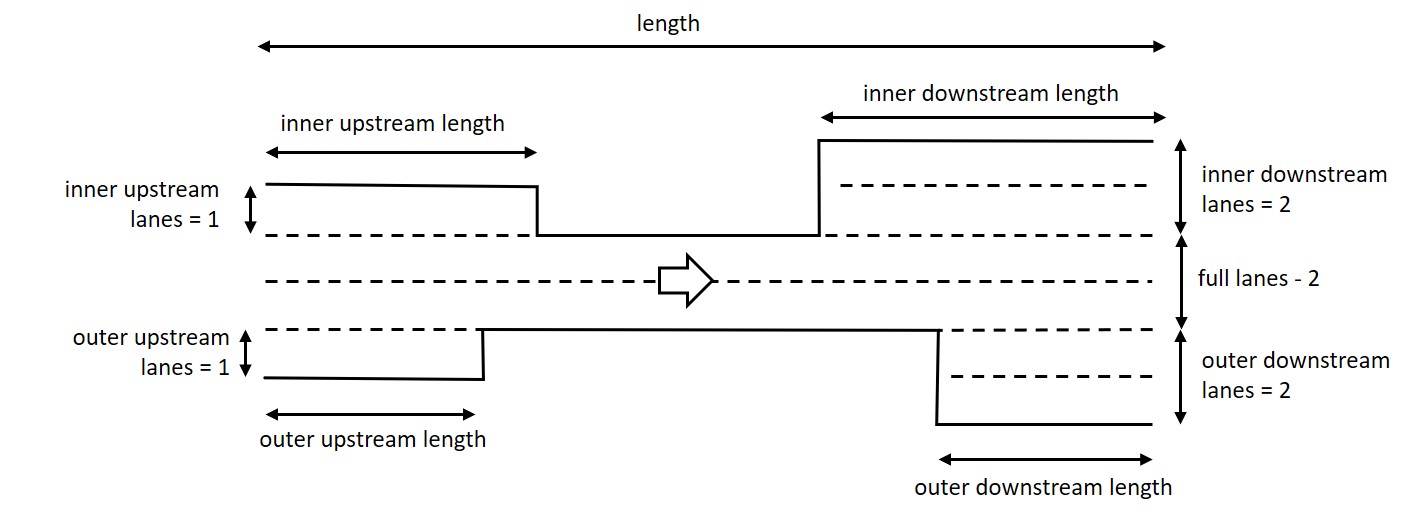}
    \caption{Generic road geometry.}
    \label{fig:link_structure}
\end{figure}

The \textit{road network} consists of an interconnection of road segments, or \textit{links}. The state of each link is managed by its \textit{model}. Figure~\ref{fig:link_structure} shows a generic link. Each link has a positive number of full-length lanes. It may also have partial-length lanes, such as turn pockets. 
The four possible partial lane structures are in the ``inner-upstream'', ``inner-downstream'', ``outer-upstream'', and``outer-downstream'' positions. Each partial lane structure is characterized by its position, number of lanes, length, and gates.

\begin{figure}[ht]
    \centering
    \includegraphics[width=\textwidth]{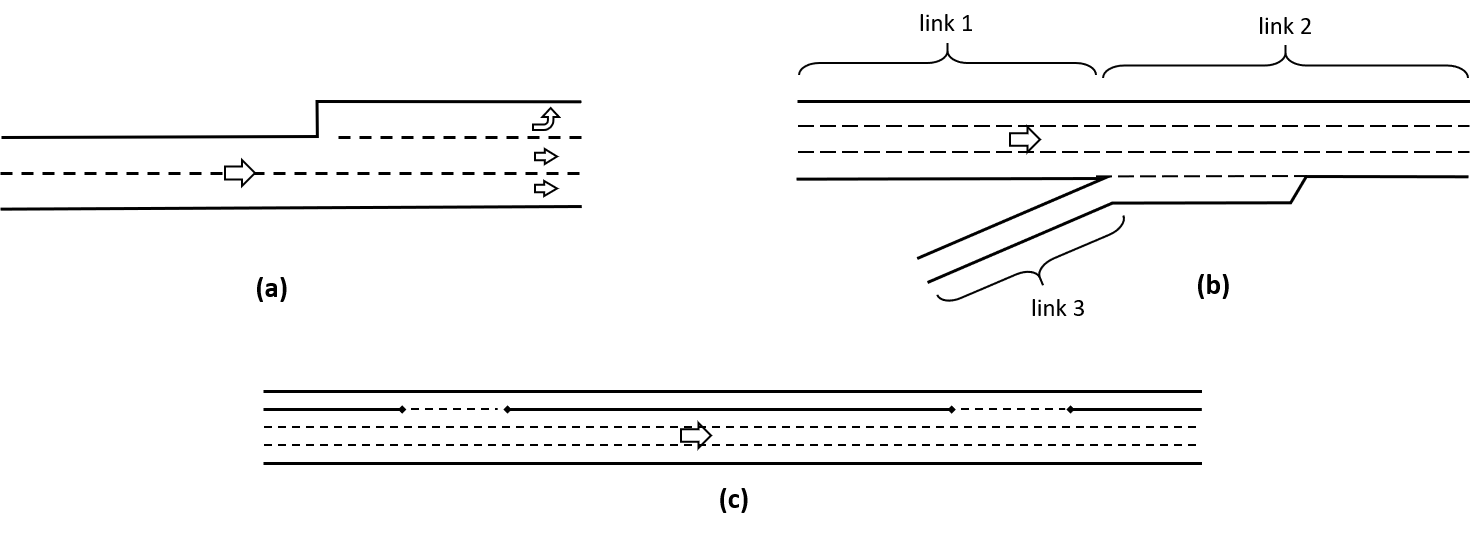}
    \caption{Example road geometries}
    \label{fig:configs}
\end{figure}

Figure~\ref{fig:configs}(a) shows an intersection approach. The turn pocket is captured as an inner-downstream partial lane. Figure~\ref{fig:configs}(b) shows an onramp merge. The merge lane is represented as an outer-upstream partial lane in link 2. In both cases, access to the partial lanes is \textit{unrestricted}. Figure~\ref{fig:configs}(c) depicts a freeway segment with a priority lane with \textit{restricted} access. The priority lane can be represented as an inner-downstream (or upstream) partial lane structure with length equal to the full length of the link. Access to the priority lane can be restricted to a series of \textit{gates}, or unrestricted. 

\subsection{Road parameters}
\begin{figure}[ht]
    \centering
    \includegraphics[width=3in]{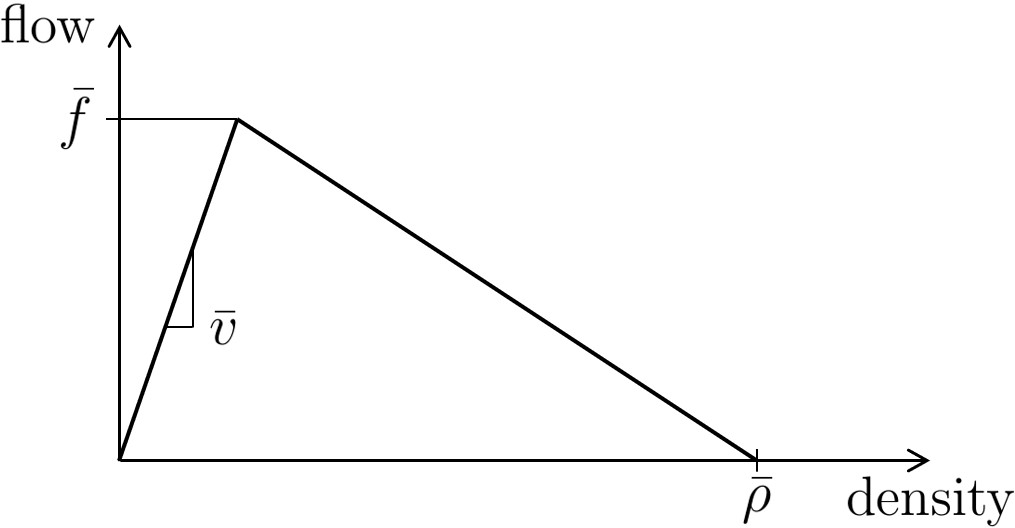}
    \caption{Required road parameters}
    \label{fig:fd}
\end{figure}

The driving characteristics of the road segments are captured by a set of road parameters. These parameters are, in general, specific to the model that manages the link. However, there are basic roadway characteristics that are common to many traffic models. These are the three parameters of the so-called triangular fundamental diagram: the road capacity ($\bar{f}$), the speed limit ($\bar{v}$), and the jam-density ($\bar{\rho}$). See Figure~\ref{fig:fd}.

It is required that the per-lane values of these three quantities be defined for each road segment. These are interpreted by the models and translated into model-specific parameters. Thus, a basic degree of physical consistency between models is achieved. The models may also require additional parameters to be defined, and these can be provided separately. Section~\ref{sec:models} describes how these quantities are interpreted by each of the three canonical models included with OTM. 

\subsection{Road connections}
Thus far we have focused on the geometry and characteristics of isolated road segments. We will now describe how these road segments are connected. The standard approach for macroscopic traffic models is to use a graph in which links are identified with edges and their interconnections are represented as vertices. This approach ignores the finer-grained interconnections between lanes, which are needed by most microscopic models. Here we use \textit{road connections} to specify the relations between links at the lane level. 

\begin{figure}[ht]
    \centering
    \includegraphics[width=0.6\textwidth]{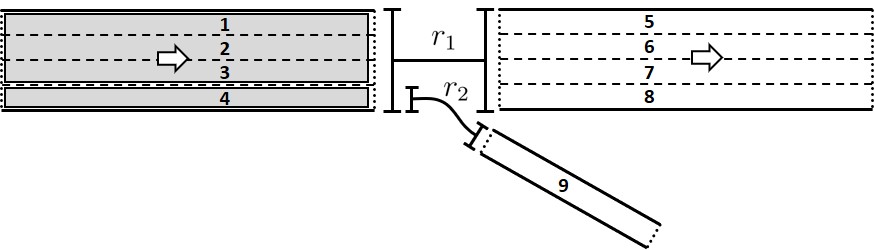}
    \caption{A freeway offramp}
    \label{fig:offramp}
\end{figure}

A road connection is a tuple with two elements: a set of upstream lanes and a set of downstream lanes. $r=(\lanesupr,\lanesdnr)$. The interpretation of a road connection is that vehicles that depart the upstream link from a lane contained in $\lanesupr$ may enter the downstream link through any lane in $\lanesdnr$. Figure~\ref{fig:offramp} shows a representation of a freeway offramp. The three links are the upstream freeway segment (lanes 1 to 4), the downstream freeway segment (lanes 5 to 8), and the offramp (lane 9). The road connections are $r_1 = (\{1,2,3,4\},\{5,6,7,8\})$ and $r_2 = (\{4\},\{9\})$.
Movements that do not follow road connections are prohibited. For example, vehicles are not allowed to exit the freeway from lane 3. The granularity of the representation can be controlled by adding or removing road connections. For example, vehicles can be made to preserve their lanes when moving from one link to the next by assigning road connections to every individual lane. However this level of detail is often not needed. 

There may be overlap between the upstream lanes of two road connections ($\lanesupr\cap\lanesuprp\neq\emptyset$), and thus a vehicle within the overlap may have turning options (e.g. lane 4 in Figure~\ref{fig:offramp}). It is required that these options lead to different links, so that split ratios specified at the link level may be translated into split ratios between road connections. 

\subsection{Lane groups}

We use the term \textit{lane group} to mean a set of adjacent lanes that share the same set of exiting road connections. Lanes in a single lane group can be expected to advance with approximately equal speed. The model that operates on the link can therefore assign to each lane group a single speed-synchonized structure. For a microscopic model this would be a lane; for a mesoscopic model, a FIFO queue; for a macropscopic model it would be an instantiation of the differential equation. The model may \textit{choose} to create a finer representation (e.g. lane-by-lane), but this is not required.

\subsection{Models and state indices}
A \textit{model} is a program that manages the state of one or more links in the simulation. The responsibilities of a model are,
\begin{enumerate}
\item to maintain the internal state of the lane groups in its links, 
\item to emit vehicle packet release requests to the simulator, in order to send vehicles to a downstream model,
\item to ensure that vehicles only leave the link along road connections that are consistent with their routing,
\item to receive incoming vehicle packets sent from upstream links, 
\item to respond to requests for information from the simulator.
\end{enumerate}

The system requires that models keep track of the type and routing information of all of the vehicles in their links. The routing information is captured by the route id, if the vehicle is of a \textit{routed} type, and by the id of the next downstream link, if the vehicle is of a \textit{probabilistic} type. Thus it is assumed that all vehicles (probabilistically routed vehicles in particular) select upon entering a link, the link to which they will proceed following that link. This assumption is in contrast to many graph-based models, in which the split ratio is only applied as vehicles flow through a multi-output node; that is, when they \textit{exit} the link. This selection of a next link, whether deterministic (provided by the route) or probabilistic (provided by split ratios) is computed by the OTM system and attached to the vehicle packets as they enter the link (see Section \ref{sec:model_interactions}). 

We use the term \textit{state index} to refer to a pair of vehicle type id and either route id or next link id, depending on whether the type is routed or probabilistic. Table~\ref{tab:state} shows an example of a lane group with four state indices. The first and second rows correspond to vehicles of type 1, the third and fourth are vehicles of type 2. If type 1 is routed, then the third column indicates that these vehicles are on routes 7 and 4. If type 2 is probabilistic then the third column indicates that these vehicles will proceed to links 23 and 34 after leaving the present link. The values of the states are listed in column four, and are in units of number of vehicles. These values are defined and managed by the model. The model represented in the table is a first order fluid model. If it were a second order fluid model, then the values of the fourth column would be two-dimensional arrays. If it were a vehicle-based model, then they would be a set of vehicle objects. 

\begin{table}[ht]
\centering
\begin{tabular}{ccc} \hline
Type & Route or Next link & Vehicles \\ \hline
1 & 7 & 11.2 \\ \hline
1 & 4 & 4.7 \\ \hline
2 & 23 & 0.3  \\ \hline
2 & 34 & 9.1 \\ \hline
\end{tabular}
\caption{Typical lane group state for a first order fluid model.}
\label{tab:state}
\end{table}

\section{Model interactions}
\label{sec:model_interactions}
Each of the individual models engaged in the simulation are responsible for managing their internal states. The interactions between these models are coordinated by the simulator using the protocol described in this section. This protocol is intended to be compatible with the Godunov scheme for partial differential equations, but also capable of handling discrete vehicles. The Godunov scheme, as illustrated by the cell-transmission model, determines the flow across the boundary between links as the minimum between what the upstream link `wants' to send (the demand) and what the downstream link can accommodate (the supply). The method can also be used to solve second-order fluid dynamical models by generalizing the definitions of demand and supply \citep{Lebacque}. Here we use a similar approach for passing \textit{flux packets} from an upstream model to a downstream model. The packet may contain any number of vehicles, along with their state, and tagged with a \textit{state index}. For example the contents of Table~\ref{tab:state} is a viable flux packet. Packets will contain either whole vehicle objects or arrays of real numbers, depending on whether the originating model is vehicle-based or fluid-based. OTM provides basic object types for vehicle-based and fluid-based state descriptions with translators for converting between them. These can be used to build new model plugins.

\begin{figure}[h!]
    \centering
    \includegraphics[scale=0.5]{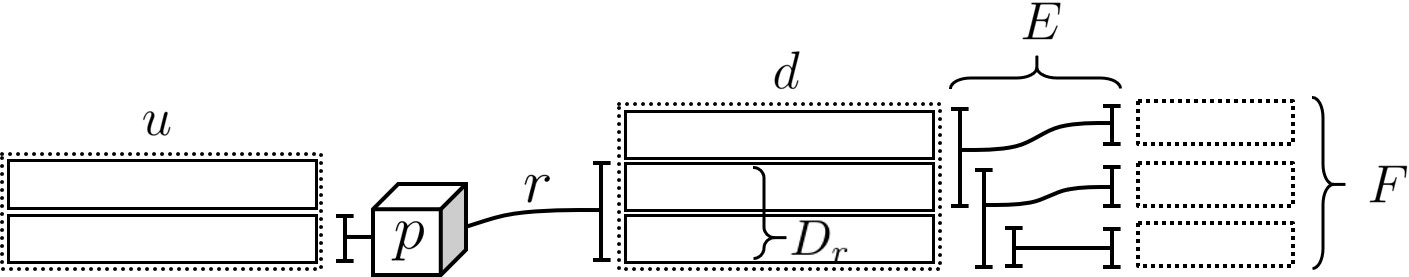}
    \caption{Links are drawn with a dotted rectangle, lane groups with a solid rectangle, and nodes with a circle.}
    \label{fig:protocolelements}
\end{figure}

The protocol is triggered when the upstream model $m_u$ requests to send a packet $p$ along a road connection $r$. Figure~\ref{fig:protocolelements} shows the elements involved. The road connection $r$ leads to a set of lane groups $D_r$ in the downstream link $d$. The set of road connections that leave link $d$ are denoted with $E$. The set of links reached by road connections in $E$ is $F$.

Upon receiving the request from the upstream model to release $p$ along $r$, the system asks the downstream model how much space the packet $p$ would occupy in link $d$, and how much space is available. This is done by invoking the following methods in the model interface, as implemented by $m_d$.

\begin{align}
\label{eq:norm}
\texttt{get\_packet\_size} &\;:\; (p,r)\rightarrow |p|\in\mathbb{R}^+ \\
\label{eq:maxpacket}
\texttt{get\_max\_packet\_size} &\;:\; (p,r)\rightarrow \bar{p}\in\mathbb{R}^+
\end{align}
These are then used to compute a scaling factor for packet $p$,
\begin{equation}
\label{eq:alpha}
\alpha = \min\left( 1 , \frac{\bar{p}}{|p|} \right)
\end{equation}
It is assumed that by multiplying the number of vehicles in the packet by $\alpha$ (uniformly over all state indices), while keeping fixed all other states and vehicle characteristics, the scaled packet ($\alpha p$) will fit in the downstream link. This is true for first order fluid models, and for the class of generic second order models described in \citet{Lebacque} (GSOM class).
If the sending model is vehicle-based, then the packet must be split into two integer-valued parts, while complying with the space constraint of Eq. (\ref{eq:alpha}). In this case, the calculation is more complicated: each state index is scaled separately by the largest amount that preserves whole vehicles while not exceeding the value of $\alpha$ from Eq. (\ref{eq:alpha}).

In the particular case of a downstream mesoscopic model with vertical queueing, $\alpha$ always evaluates to 1, since all packets can be accepted without restriction. For second-order fluid models, such as those in the GSOM class, the maximum packet size depends on the composition of the packet, in addition to the state of the downstream link. This is the reason for including $p$ as a parameter in Eq. (\ref{eq:maxpacket}). 

It should be noted that it is not necessary that all models share the same `norm'; that is, they need not agree on the size of a packet. They must however return a value $|p|$ such that $\alpha p$ (as defined above) can be accommodated.

Next, before delivering $\alpha p$ to the downstream model, the system determines the next link for probabilistically routed vehicles. This is done by sampling the split ratios assigned to links in $F$ in Figure~\ref{fig:protocolelements}. If the upstream model is fluid-based, then $\alpha p$ is divided into as many smaller packets as non-zero split ratios exist for the given vehicle type. If it is vehicle-based, then the each vehicle is assigned a next link with probabilities correseponding to the split ratios. The resulting collection of vehicles packets is then passed to the downstream model using the method,
\begin{equation}
\label{eq:sendpacket}
\texttt{send\_packets} \;:\; (\mathcal{P},r)\rightarrow \text{void}
\end{equation}
Here $\mathcal{P}$ is the collection of packets being sent, and $r$ is the road connection along which they travel. Upon receiveing the packets, the downstream model must distribute them amongst the lane groups in $D_r$. On possible strategy for doing this is to split each packet in $\mathcal{P}$ into $|D_r|$ equal packets. Another is to split them in a way that seeks equilibrium between the lane groups. OTM provides implementations of these two strategies that can be used by new model plugins. 

After lane group assignment, the final step of the process is to translate all of the packets into the native representation of the receiving model, and incorporate them into the state. This is done by the downstream model using vehicle classes and translators provided by the program.

\section{Simultaneous requests}
\label{sec:nodemodel}
The previous section described the interaction between an upstream and a downstream model when a single vehicle packet is sent between the two. In general however, packets may be sent simultaneously through a multi-input/multi-output connection. This is the case, for example, with discrete-time fluid models where all lane groups emit packets at every time step. In this section we describe the extension of the model coordinator for this case. This is known in the transportation literature as a ``node model''. The purpose of the node model is to scale the packets that travel through a MIMO node such that they can be acommodated by the downstream lane groups, while satisfying some criterion such as maximizing flow. 
We describe here the particular node model that has been implemented in OTM, which is an adaptation of the method of \citet{Wright} to a network representation consisting of lane groups and road connections, instead of nodes and links. 

\begin{figure}[!ht]
    \centering
    \includegraphics[width=0.9\textwidth]{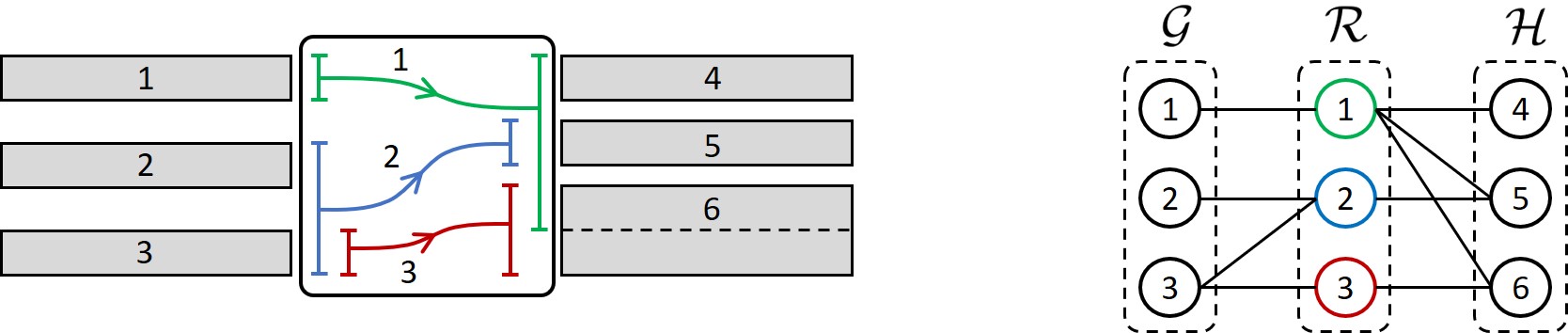}
    \caption{Node model.}
    \label{fig:nodemodel}
\end{figure}

Figure~\ref{fig:nodemodel} provides an illustration. A set of packets is requested to be sent from upstream lane groups in the set $\GG$, through road connections in the set $\RR$, to downstream lane groups in the set $\HH$. The ``node'' in the figure has three incoming, or upstream, lane groups \{1, 2, 3\}, which connect via three road connections \{1, 2, 3\} to three downstream lane groups \{4, 5, 6\}. The figure also shows a graph representation of the connectivity between these elements. This graph is encoded in ``upstream'' and ``downstream'' sets for each element. The downstream set of lane group 1 is road connection 1, the upstream set of lane group 6 is road connections 1 and 3, etc. These sets are denoted with $\Dg$, $\Ur$, $\Dr$, and $\Uh$ for the upstream and downstream sets of elements in $\GG$, $\RR$, and $\HH$. 

\begin{center}
\begin{BVerbatim}
while true {
    (NM 0)
    break if (stopping criterion)
    (NM 1) through (NM 6)
}
\end{BVerbatim}
\end{center}
The execution of the node model is shown above. With each iteration, the upstream models send a portion of their packets. The portion is limitted either by upstream demand or downstream supply, and hence, as with \citet{Wright}, the algorithm is guaranteed to complete within a number of iterations not exceeding the number of lane groups. We confine the description to the simpler case, where the total supply in downstream lane groups are independent of the composition of the packets being sent to those lanes groups. This case covers first-order fluid models and most vehicle-based models. This more general case has been developed by \citet{Wrightnodemodel} for graph representations, and will be developed for OTM in a future publication. 

\subsection{Node model steps}

\begin{enumerate}[label=\texttt{(NM \arabic*)}]
\setcounter{enumi}{-1}
\item $\demandgr\in\mathbb{R}^+$ is the size of the packet emitted by lane group $\alg\in \GG$ along road connection $r\in\RR$. This is obtained, as in the SISO case, by invoking Eq. (\ref{eq:norm}) on the unique receiving model. For each lane group $\alg$, we construct the set of road connections with positive demand:
\begin{align}
\Dgp &= \{ r\in\Dg : \demandgr>0 \} & \forall \alg\in\GG
\end{align}
Boolean variables $\blockedg$, $\blockedr$, and $\blockedh$ are used to indicate that a lane group or road connection is ``blocked''. A downstream lane group $h$ is blocked if it has zero supply $\supplyh$. 
\begin{align}
\label{eq:blockedh}
\blockedh &= [ \supplyh==0 ]  & \forall h\in\HH 
\end{align}{}
The supply $\supplyh$ is obtained from Eq.~(\ref{eq:maxpacket}). This function is expected to return zero whenever lane group $h$ is full, and thus prevent vehicles from entering. A road connection is blocked only if \textit{all} of its downstream lane groups are blocked. This reflects the assumption that packets traveling along road connection $r$ may be placed in any of the lane groups in $\Dr$. 
\begin{align}
\label{eq:blockedr}
\blockedr &=  [ \blockedh ]_{\forall h\in\Dr} & \forall r\in\RR  
\end{align}
The variable $\blockedg$ is true whenever lane group $\alg$ is either \textit{blocked} or \textit{empty}. An upstream lane group is \textit{blocked} whenever \textit{any} of its exiting road connections is blocked, provided there is demand for that road connection (Eq.~\ref{eq:blockedg}). This reflects the FIFO assumption within the lane group: a single vehicle moving along a blocked road connection will block the entire upstream lane group. A lane group is \textit{empty} if its demand has been delivered, and therefore the set $\Dgp$ is empty.
\begin{align}
\label{eq:blockedg}
\blockedg &=  [ \blockedr ]_{\exists r\in\Dgp}  \; \lor \; \left[\Dgp=\emptyset\right] & \forall \alg\in\GG
\end{align}
\end{enumerate}

\begin{enumerate}[label=\texttt{(stopping criterion)}]
\item The computation halts if all upstream lane groups are blocked or empty.
\begin{equation}
\text{stop if } [ \blockedg ]_{\forall \alg\in\GG}
\end{equation}
\end{enumerate}

\begin{enumerate}[label=\texttt{(NM \arabic*)}]
\item For every $r\in\RR$: Calculate the total demand on each road connection by aggregating over incoming lane groups.  
\begin{equation}
\demandr = \sum_{\alg\in\Ur} \demandgr
\end{equation}       
Compute the downstream supply available to each road connection, disregarding the demands from other road connections. This calculation involves $\laneproprh\in(0,1]$, the portion of lane group $h$ that is accessible to road connection $r$. In Figure~\ref{fig:nodemodel}, $\lambda^1_6=0.5$ since road connection 1 has access to only one of the two lanes in lane group 6. 
\begin{align}
\supplyr &= \sum_{h\in \Dr} \laneproprh \supplyh
\end{align}
Compute the apportionment of the demand on connector $r$ to each of its downstream lane groups $h$. This is done in proportion to the available supply.
\begin{align}
\demandproprh &= \left\{
\begin{tabular}{ll}
0 & if $\supplyr==0$  \\
$\laneproprh \supplyh/\supplyr$ & otherwise \\
\end{tabular}
\right. & \forall \;h\in\Dr
\end{align}

\item For every $h\in\HH$: Compute the total demand on downtream lane group $h$ as the sum of demands on its incoming road connections scaled by the apportionment factor (Eq.~\ref{eq:demandh}). 
$\reductionh$ is the percent excess demand on lane group $h$ (Eq.~\ref{eq:reductionh}).
\begin{align}{}
\label{eq:demandh}
\demandh &= \sum_{r\in\Uh} \demandproprh \demandr \\
\label{eq:reductionh}
\reductionh  &= \max \left( 0, 1-\frac{\supplyh}{\demandh}    \right)
\end{align}

\item For every $r\in\RR$: Propagate the excess demand factors to the road connectors. The demand on each road connector must be reduced by the worst-case factor. Note that this formula will yield $\reductionr=1$ for a blocked road connection, since with $\supplyh=0$ we have $\reductionh=1$ for all of its downstream lane groups. 
\begin{equation}
\reductionr = \sum_{h\in\Dr} \demandproprh \reductionh
\end{equation}

\item For every $g\in\GG$: Compute the reduction factor for each upstream lane group. Only road connections with positive demand ($r\in\Dgp$) are considered in this calculation. The $\blockedg$ case is included to account for the situation that $g$ is empty, and therefore the $\max$ is undefined. If $g$ is blocked, then both formulas evaluate to 1. 
\begin{align}
\reductiong &= \left\{
\begin{tabular}{cr}
1 & $\blockedg$ \\
$\max_{r\in\Dgp} ( \reductionr)$ & $!\blockedg$
\end{tabular} \right.
\end{align}
Compute the demand that advances along each road connection exiting lane group $g$ (Eq.~\ref{eq:deltagr}), and update the demand that remains (Eq.~\ref{eq:demandgr}). $\forall\;r\in\Dgp$:
\begin{align}
\label{eq:deltagr}
\deltagr &= \demandgr (1-\reductiong)  \\
\label{eq:demandgr}
\demandgr &\leftarrow \reductiong\demandgr 
\end{align}

\item For every $r\in\RR$: Calculate the portion of the advancing flow on each road connection. 
\begin{align}
\deltar &= \sum_{g\in\Ur} \deltagr 
\end{align}

\item For every $h\in\HH$: Collect the flow entering each downstream lane group and reduce the downstream supplies. 
\begin{align}
\deltah &= \sum_{r\in\Uh} \frac{1-\reductionh}{1-\reductionr} \demandproprh \deltar  \\
\supplyh &\leftarrow \supplyh - \deltah
\end{align}
\end{enumerate}

Once the algorithm converges, the scaling factors that are the output of the node model can be computed from the total amount of each packet that has been sent.

\section{Models}
\label{sec:models}
Models are implemented in OTM as plugins, that is, as relatively small pieces of code that define the functionality specified by the modeling interface. This interface includes methods such as Eqs.~(\ref{eq:norm}), (\ref{eq:maxpacket}) and (\ref{eq:sendpacket}), as well as other methods that are needed to track performance measures. Additionally, the model must provide methods for advancing its internal state in time. That is, it must provide the longitudinal and lateral dynamics. All other functionality -- routing, traffic control, performance metrics, execution, and process parallelization -- are managed by OTM. The main requirement placed on the model is that it must only release vehicle packets along road connections that are consistent with the routing behavior of the vehicles. This implies that vehicles must move laterally within a link (from one lane group to another) in order to reach a valid exiting road connection. In short, the model must implement a lane-changing strategy.

This section describes implementations of the three canonical examples for microscopic, mesoscopic, and macroscopic modeling; respectively, Newell's simplified car following model, the two-queue model, and the cell-transmission model.

\subsection{Newell's car-following model}
\label{sec:carfollowing}
The model described here is a discrete-time version of \citet{Newell}. Each lane group contains a single first-in-first-out queue of vehicles, numbered starting with the downstream-most vehicle. The position of the $i$th vehicle at time $t$, $x_i(t)$, is related to the position of the $(i-1)$th vehicle at a previous time,
\begin{equation}
\label{eq:xit}
x_{i}(t+\Delta t) = x_{i}(t) + \max\left(\: 0 \:,\: \min \left(\:  \delta v^i \;,\; h_i(t) - \delta w^i \;,\; h_i(t) \delta\bar{f}^i\:\right)   \right)
\end{equation}
Here $\Delta t$ is model's simulation time step; $\delta v^i$, $\delta w^i$, and $\delta \bar{f}^i$ are sampled from distributions whose means correspond to the left and right slopes of the fundamental diagram, and the capacity (see Figure~\ref{fig:fd}). 
\begin{align}
\delta v^i &\sim \mathcal{N}(v_f \Delta t ,\vardxvi) \\
\delta w^i &\sim \mathcal{N}(w \Delta t ,\vardxwi) \\
\delta \bar{f}^i &\sim \mathcal{N}(\bar{f} \Delta t ,\sigma^f)
\end{align}
$\vardxvi$, $\vardxwi$, and $\sigma^f$ are user-defined standard deviations. The first term in Eq. (\ref{eq:xit}) applies to lead vehicles that travel with free-flow speed $v_F$. The second term involves $h_i(t)$, the headway of the $i$th vehicle at time $t$. This is defined, for all vehicles except the first vehicle in the lane group, as the distance to its leader: $h_i(t)=|x_i(t)-x_{i-1}(t)|$. The first vehicle however does not have a leader in its lane group, and hence it is possible that its leader belongs to a different model. Computing the headway for the first vehicle may therefore involve requesting the position of the upstream-most vehicle in the lead vehicle's next link. This is done by calling a method provided by the modeling interface.
\begin{equation}
\texttt{get\_distance\_to\_last\_vehicle}: \; (r) \rightarrow \eta\in\mathbb{R}^+
\end{equation}
This measures the distance from the upstream boundary of the link accessed by road connection $r$ to the nearest vehicle in any of the lane groups accessible from $r$. 

The third term in Eq.~(\ref{eq:xit}) imposes a capacity constraint. 

The lane change model that has been implemented for this model is simple. Vehicles that enter a link are placed directly into a lane group that connects to their target road connection, irrespective of whether the road connection they enter by actually connects to that lane group. In other words they change lanes immediately and without obstruction into their target lane group. If the target lane group is already full, then the vehicle is placed in a buffer attached to that lane group, and enters as soon as space becomes available.

\subsection{Two-queue model}
\label{sec:queueing}
\begin{figure}[!ht]
    \centering
    \includegraphics[width=4in]{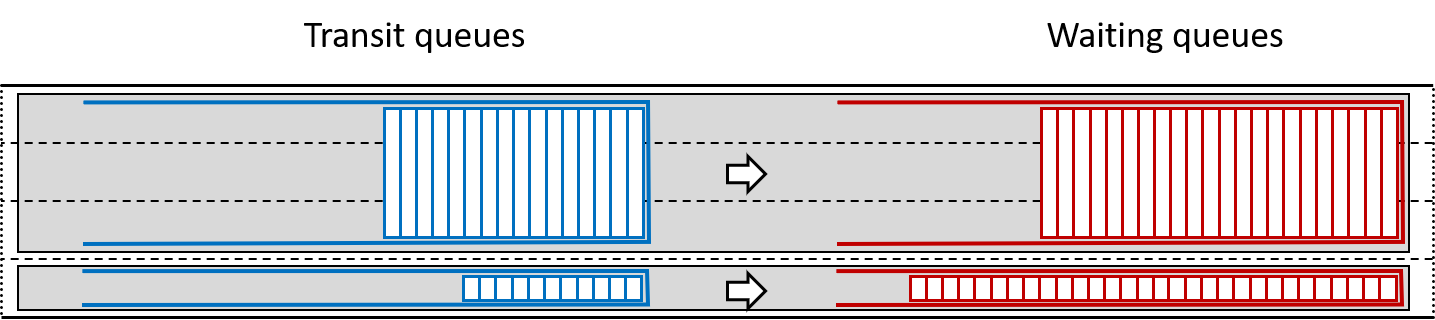}
    \caption{Queueing model.}
    \label{fig:queueing}
\end{figure}

This mesoscopic model is similar to the one reported in \citet{Varaiya}. Each lane group is equipped with two queues.  The first is a \textit{transit} queue, which delays every vehicle entering the lane group by the free-flow travel time. After leaving the transit queue, the vehicle enters the \textit{waiting} queue, which is a FIFO queue, serviced by a Poisson process. See Figure~\ref{fig:queueing}. 

This model makes use of the three supply-side parameters of Figure~\ref{fig:fd}. The speed limit $\bar{v}$ is used to calculate the free-flow travel time. The capacity $\bar{f}$ is the average discharge rate for the waiting queue. The jam density $\bar{\rho}$ is used to compute the available supply for incoming vehicle packets. These packets are rejected whenever the combined number of vehicles in the transit and waiting queues reaches the maximum value. The lane changing model for the queuing model is identical to that of the car-following model: an arriving vehicle is placed immediately into its target lane group unless it is full, in which case the vehicle is held in a buffer until space becomes available.

\subsection{Cell-transmission model}
\label{sec:fluid}
\begin{figure}[h!]
    \centering
    \includegraphics[width=0.9\textwidth]{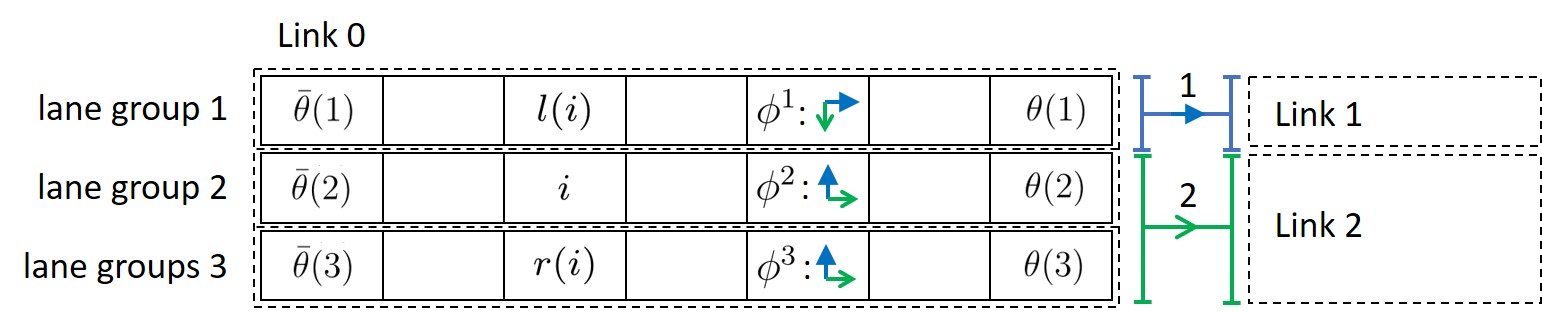}
    \caption{Lane changing model for the CTM.}
    \label{fig:ctm}
\end{figure}

The macroscopic model included with OTM is an adaptation of the cell transmission model (CTM) of \citet{Daganzo1994}, with a lane change strategy that is similar to that of \citet{Laval}. Lane groups are divided into \textit{cells}. Figure~\ref{fig:ctm} shows a single link with three lane groups. All of the cells in a link are of equal size, which is computed as the largest that yields an integer number of cells, without exceeding a user-defined \textit{maximum cell size}. The simulation time step $\Delta t$ must comply with the Courant–Friedrichs–Lewy condition, that no cell can be traversed in one time step by a vehicle traveling at maximum speed. 

Each cell $i$ has up to two lateral neighbors in adjacent lane groups; the inner and outer lane groups ($in(i)$ and $out(i)$). The state of a cell $i$ consists of the number of vehicles per state index $s$: $n^i_s$ (omitting the time index for convenience). 

$\linkstates$ is the set of all states that can use link $\alink$. The link shown in Figure~\ref{fig:ctm} carries two states: one represented by a solid blue arrow, which is headed for link 1, and another represented by an open green arrow, which is headed for link 2. 
Each lane group carries a map $\rho^g$ that returns, for each state, the road connection $r$ that it must use to reach its next link.
\begin{equation}
\rho^g : \linkstates \rightarrow \mathcal{R} 
\end{equation}
In Figure~\ref{fig:ctm} we have that $\rho^1$ evaluated on the blue state returns road connection 1, while $\rho^2$ and $\rho^3$ evaluated on the green state return road connection 2. It is a requirement on the network configuration that this map should return at most one road connection. That is, for each lane group $g$, the road connections that leave $g$ must all lead to different links. 

We define the \textit{target lane group set} for a state $s$ within link $\alink$ as those from which its next link can be reached. In the figure, the target lane group set for the blue state is \{1\}, and for the green state it is $\{2,3\}$. Each lane group also carries a map $\phi^g$ indicating the direction in which a state must change lanes in order to reach its target lane group set.
\begin{equation}
\phi^g : \mathcal{S}^{\ell(g)} \rightarrow \{in,out,none\}
\end{equation}
Here $\ell(g)$ is the link that contains lane group $g$, and hence $\mathcal{S}^{\ell(g)}$ is the set of states that use $g$. $\phi^g$(s) returns $in$, $out$, or $none$ depending on whether the state must change lanes inward, outward, or not at all to reach its target road connection. Figure~\ref{fig:ctm} illustrates this map. 

\subsection{Dynamics}

Every lane group managed by the fluid-dynamical model emits vehicle packets every $\Delta t$. These packets are assessed by the OTM node model, which involves calls to the supply calculation function. Following this, every lane group receives a set of vehicle packets from its incoming road connections, and proceeds to update its internal state. These three steps -- vehicle packet generation, supply calculation, and state update -- are described below.

\subsubsection{Vehicle packet generation.}
\label{sec:fluxpacket}
This stage consists of the five internal steps. First, the number of vehicles that change lanes during the upcoming time step is computed with Eqs. (\ref{eq:fp1}) through (\ref{eq:fp3}), and used to compute an intermediate state with Eq. (\ref{eq:intermediatestate}). The intermediate state is then used to evaluate the longitudinal demand, in step Eqs. (\ref{eq:demandgrs}).

\begin{figure}[ht]
    \centering
    \includegraphics[width=3in]{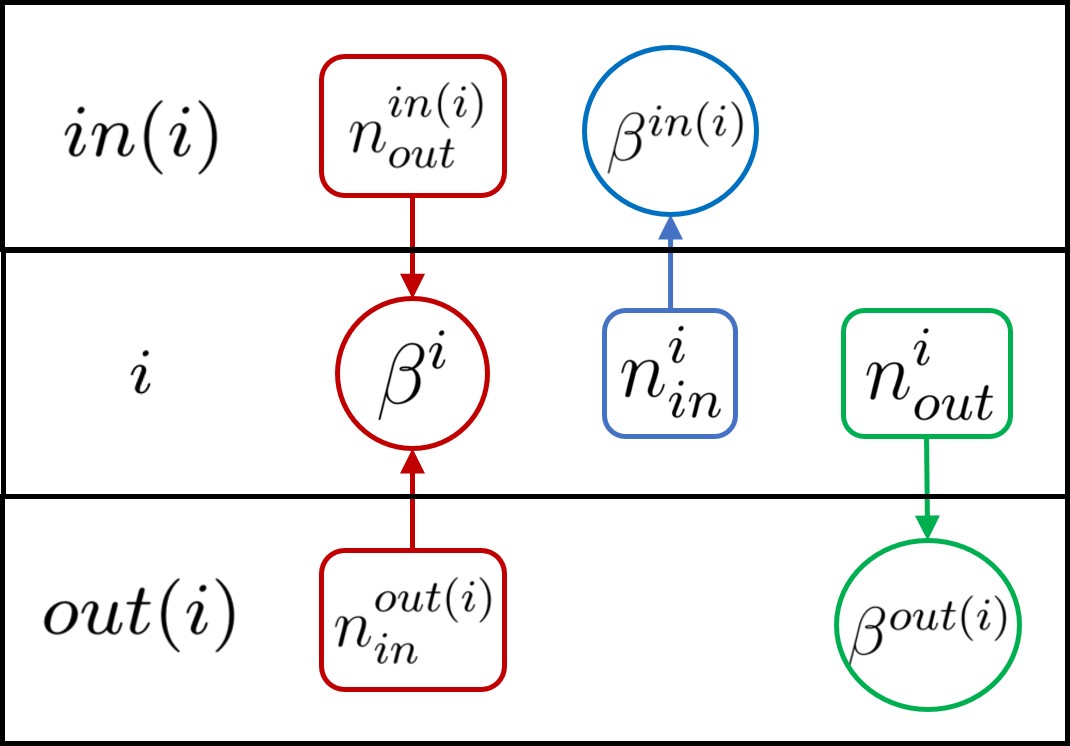}
    \caption{Lane change model.}
    \label{fig:lanechange}
\end{figure}

\begin{enumerate}[label=\texttt{[FP \arabic*]}]
\item Compute, for each cell, the total number of vehicles performing each of the lane change maneuvers. 
\begin{align}
\label{eq:fp1}
n^i_\mu &= \sum_{\{s: \phi^{g(i)}(s)=\mu\}}n^i_s & \mu\in\{in,out,stay\} 
\end{align}
Here, $g(i)$ is the lane group for cell $i$, and therefore $\{s: \phi^{g(i)}(s)=\mu\}$ is the set of states that perform maneuver $\mu$ in cell $i$.
\item Compute for all cells $i$ the total number of vehicles in the cell.
\begin{align}
\label{eq:fp2}
n^i &= n^i_{in} + n^i_{out} + n^i_{stay}= \sum_{s} n_s^i 	& \forall \text{ cells } i 
\end{align}
\item Compute the scaling factor for all cells. $\xi^i\in[0,1]$ is a supply apportionment factor for lane change movements, analogous to $w$ for longitudinal movements. Hence the space available in cell $i$ for vehicles changing lanes is $\xi^i(\bar{n}^i-n^i)$, where $\bar{n}^i$ is the maximum occupancy. As illustrated in Figure~\ref{fig:lanechange}, the total number of vehicles entering cell $i$ is $n^{out(i)}_{in} + n^{in(i)}_{out}$, and the scaling factor $\beta^i$ is,
\begin{equation}
\label{eq:fp3}
\beta^i = \min \left(\: 1 \;,\; \frac{\xi^i ( \bar{n}^i - n^i ) }{n^{out(i)}_{in} + n^{in(i)}_{out}} \: \right) 
\end{equation}
\item Apply the scaling factors to the lane change demands and execute the maneuvers. This results in an intermediate state $\hat{n}^i_{s}$ which represents the number of vehicles for state $s$ after the lane changes have completed, but before vehicles have moved forward. 
\begin{align}
\label{eq:intermediatestate}
\hat{n}^i_{s} = &\left\{
\begin{tabular}{cl}
$\left(1-\beta^{in(i)}\right)\:n^i_s$ & if $\phi^{i}(s)=in$ \\
$\left(1-\beta^{out(i)}\right)\:n^i_s$ & if $\phi^{i}(s)=out$ \\
$n^i_s$  & if $\phi^{i}(s)=none$ \\
\end{tabular} \right\} \\
\nonumber
&+\; \beta^i\: \left\{
\begin{tabular}{cl}
$n^{out(i)}_s$ & if  $\phi^{out(i)}(s)=in$ \\
0  & otherwise\\
\end{tabular} \right\} \\
\nonumber
&+\; \beta^i\: \left\{
\begin{tabular}{cl}
$n^{in(i)}_s$ & if  $\phi^{in(i)}(s)=out$ \\
0  & otherwise\\
\end{tabular} \right\}
\end{align}
\item Compute the demand generated by each lane group on each of its outgoing road connections. This demand depends on the number of vehicles in its downstream-most cell. Denote with $\theta(g)$ a map that returns the downstream-most cell for lane group $g$ (see Figure~\ref{fig:ctm}).
\begin{equation}
\label{eq:demandgrs}
\demandgrs = \left\{
\begin{tabular}{ll}
$\min(v^{\theta(g)} \: n^{\theta(g)}_x,\bar{f}^{\theta(g)})$ & if $\rho^g(s)=r$ \\
0 & otherwise 
\end{tabular} \right.
\end{equation}
Here $\demandgrs$ is the demand for state $s$ generated by lane group $g$ on road connection $r$. $\demandgrs$ is zero if $s$ does not use $r$. Otherwise it is as prescribed by the CTM. $v^{\theta(g)}$ and $\bar{f}^{\theta(g)}$ are the normalized free-flow speed and capacity for the downstream-most cell of lane group $g$. The set of $\demandgrs$ values over all states $s$ constitutes the flux packet that lane group $g$ emits on road connection $r$.
\end{enumerate}

\subsubsection{Supply calculation.}
\label{sec:space}
The space available in lane group $g$ to receive packets from upstream road connections is a function of the state in its upstream-most cell. Denote this cell with $\bar\theta(g)$ (shown in Figure~\ref{fig:ctm}).
\begin{equation}
s^g = w^{\bar\theta(g)} ( \bar{n}^{\bar\theta(g)}  - n^{\bar\theta(g)} )
\end{equation}
Here $\bar{n}^{\bar\theta(g)}$ and $w^{\bar\theta(g)}$ are parameters of the upstream-most cell representing the jam density and congestion wave propagation speed.

\subsubsection{State update.}
\label{sec:stateupdate}
Following the completion of the node model, reduced packets are passed between the lane groups. These are then translated into the macroscopic representation, and consolidated into two packets per lane groups -- one entering and one leaving. The content of the upstream and downstream packets for lane group $g$ is the total number of vehicles per state $s$: $n^{g,up}_s$ and $n^{g,dn}_s$. The the densities per cell and state can then be updated using the law of conservation of vehicles. The incoming flow for the upstream-most cell is $n^{g,up}_s$, the outgoing flow for the downstream-most cell is $n^{g,dn}_s$. The flows across internal cell boundaries are given by the standard CTM formulas. 

All models must respond to simulator requests for the position of the upstream-most vehicle in a lane group. The CTM does this based on the supply function of the upstream-most cell in the lane group. By assuming that density concentrates toward the downstream part of the cell, the distance from the upstream edge to the last vehicle is $\ell(\bar{\rho}-\rho)/\bar{\rho}$, where $\ell$ is the length of the cell, $\bar{\rho}$ is the jam density, and $\rho$ is the current density.

\section{Control elements}
\label{sec:control}
We have thus far described the simulation of the hybrid open-loop traffic dynamics. We now describe how this dynamics can be influenced by a feedback algorithm, or a \textit{controller}. Controllers are implemented in OTM as plugins, similarly to models. They do not interact directly with the traffic models, but instead through \textit{sensors} and \textit{actuators}. A sensor is any element that extracts information from the models and provides it to the controller. An actuator delivers the control command from the controller to the models. OTM provides a number of sensors and actuators that can be used by the modeler to construct the control infrastructure. Sensors and actuators are model agnostic: they operate using methods from the modeling interface, and are not concerned with the implementation details of those methods. Thus, the traffic models used in a scenario can be changed without modifying the control structures. 

Each controller can register with one or more sensors and one or more actuators. An actuator however can only be assigned to a single controller. Each controller is prescribed a time step by the modeler. This time step need not be equal to (or a multiple of) the time steps of the underlying models. At every time step, the controller will read the measurements from its sensors, update its commands, and then transmit the commands to its actuators. Sensors and actuators have their own time steps, and again these are independent from the time steps of the controllers and models. Sensors use interface methods to extract information from the models at each sensor time step. For example, the fixed local sensor method uses the \texttt{get\_total\_vehicles\_in\_lanegroup(lg)}. 
OTM currently provides the following basic sensors,
\begin{itemize}
\item Fixed local sensor. This sensor is attached to the pavement, and can be used to extract density, flow, and speed information for a particular cross section. 
\item Fixed lane group sensor. This sensor is attached to a lane group, and can be used to obtain the total number of vehicles and their speeds. 
\item Probe. This sensor is attached to a vehicle and can be used to measure its speed and local environment.
\end{itemize}
The functionality required for the probe sensor is implemented in OTM's basic vehicle class. This class can be extended for use in any vehicle-based model. When a probe vehicle enters a fluid-based model, OTM automatically creates a `'virtual vehicle', which it tracks through the fluid network using the local speed provided by the model. 

In addition to these, there are also methods for querying static information about the scenario. This includes information about past and future demands and split ratios, controller states, and geometric information of the links and lane groups. 

These are the actuators that are currently available in OTM:
\begin{itemize}
\item Road connection blocking actuator. This actuator can be used to open and close road connections, and thus to mimick traffic signals. 
\item Variable speed limit. This actuator is used to change the speed limit of the link and its lane groups. 
\item Router. Used to change the route of a routed vehicle type.
\item Demand modifier. Alter the profile of demand intensity for a source and vehicle type.
\item Split ratio actuator. Alter the profile of split ratios for a given vehicle type at a junction.
\end{itemize}
In contrast to sensors, the implementation of these actuators does not make use of the modeling interface, but rather act by modifying the inputs to the models (e.g. road parameters).

\section{Experiments}
\label{sec:experiments}
\begin{figure}[ht]
    \centering
    \includegraphics[scale=0.5]{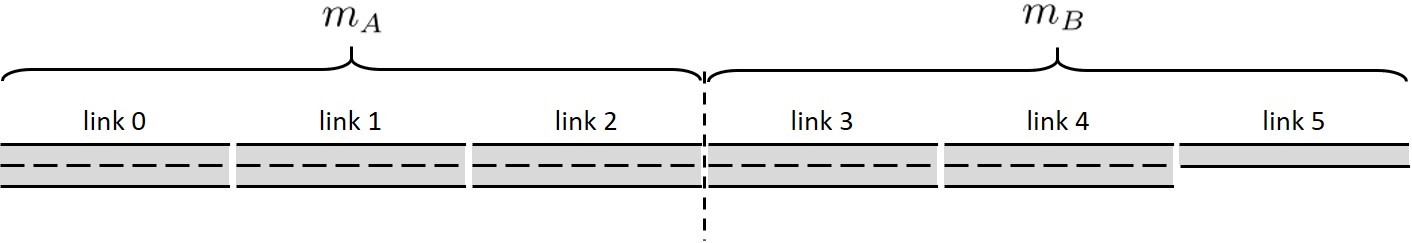}
    \caption{Experimental setup.}
    \label{fig:exp}
\end{figure}

Here we apply the OTM framework to a simple linear network with two models $m_A$ and $m_B$. This is meant only to demonstrate the translation of vehicle states across a modeling interface. It does not exercise other features of the program, such as vehicle types, lane changing, the node model, and control structures. These are left for a future publication.  
Figure~\ref{fig:exp} shows the setup. The links are all of length 500 meters. They share per-lane characteristics: the capacity is 1,000 veh/hr/lane, jam density is 100 veh/km/lane, and the free-flow speed is 100 km/hour. Links 0 through 4 have two lanes, and hence their flow and holding capacities are 2,000 veh/hr and 100 vehicles respectively. Link 5 has half of these values (1,000 veh/hour and 50 vehicles).

Links 0, 1, and 2 are managed by $m_A$; links 3, 4, and 5 by $m_B$.  A source of 1500 veh/hr is applied upstream of link 0. This causes congestion to accummulate at link 5. This congestion propagates upstream over the modeling interface, and dissipates after the demand is removed. 

\begin{figure}[ht]
    \centering
    \includegraphics[width=0.9\textwidth]{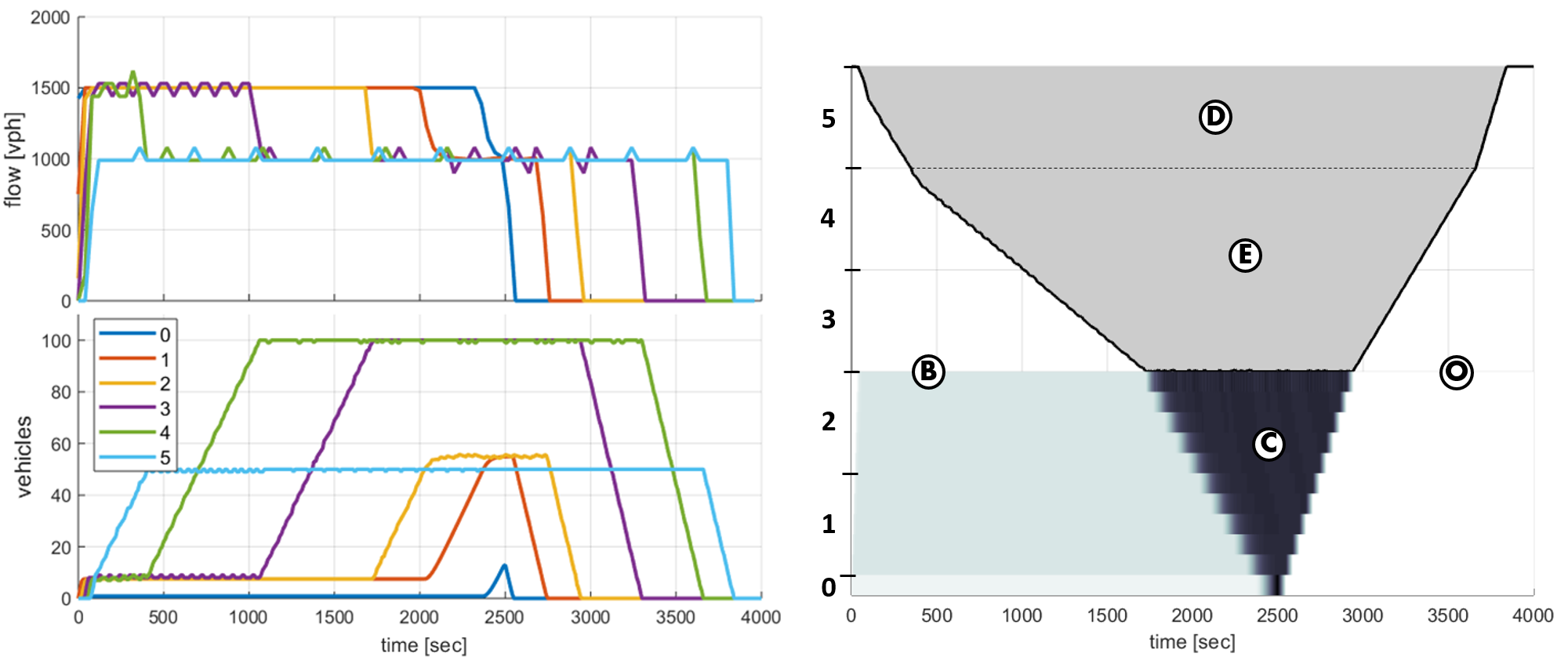}
    \caption{$m_A$=macro, $m_B$=meso.}
    \label{fig:macromeso}
\end{figure}

Figure~\ref{fig:macromeso} shows the result when $m_A$ is a macroscopic model and $m_B$ is mesoscopic.  Vehicles immediately begin to queue in link 5. The number of vehicles in link 5 reaches 50 after about 400 seconds, and then the queue spills into link 4. The queue reaches the model boundary at around 1,750 seconds and propagates into the macroscopic model. The queueing density and speed in the macroscopic portion are 55 vehicles and 9.09 km/hr, according to the hydrodynamic theory. Notice that the congested density in the fluid model is significantly lower than in the mesoscopic model (55 vehicle versus 100 vehicles). As a consequence, the speed of propagation of congestion is higher. After the demand is removed (at 2,500 second) the congestion dissipates, again at a faster pace in the macroscopic model than in the mesoscopic model. 

\begin{figure}[ht]
    \centering
    \includegraphics[width=0.9\textwidth]{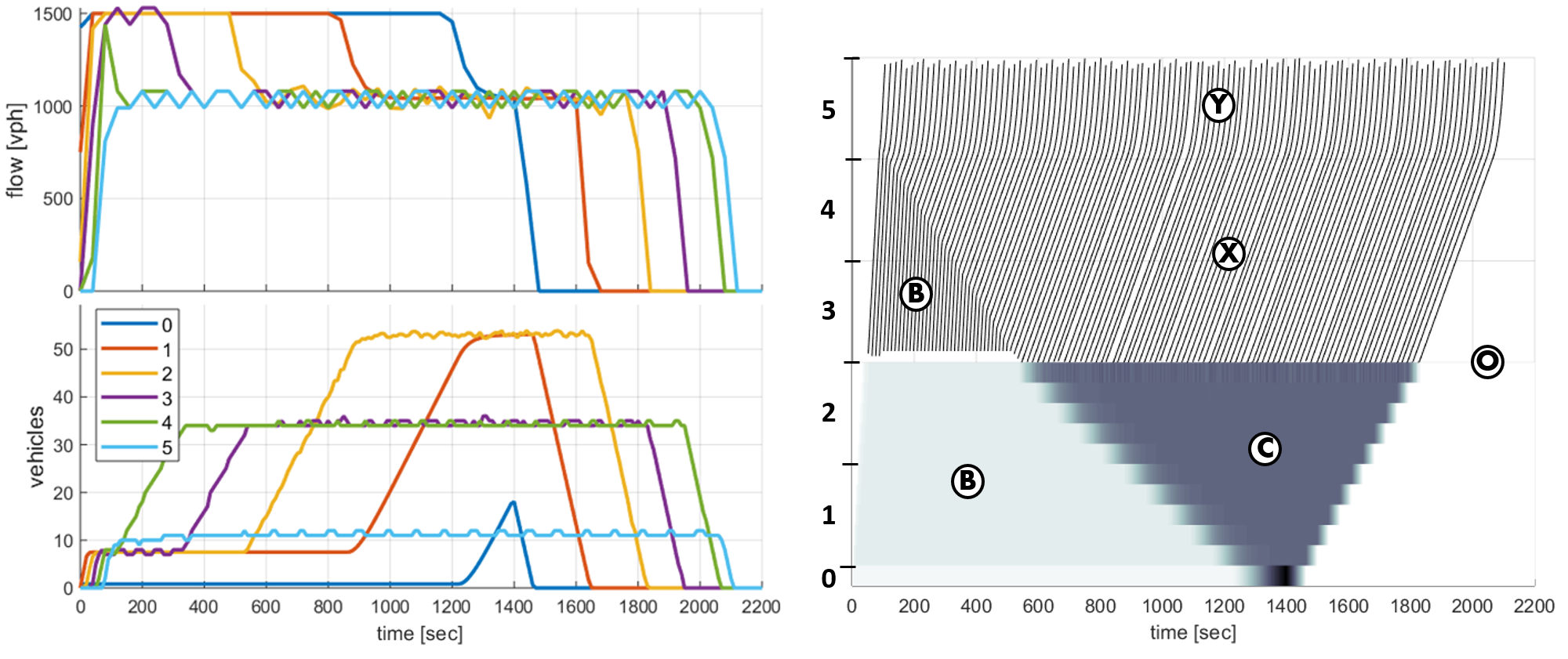}
    \caption{$m_A$=macro, $m_B$=micro.}
    \label{fig:macromicro}
\end{figure}

Figure~\ref{fig:macromicro} shows the result when $m_A$ is macroscopic and $m_B$ is microscopic. For the microscopic model, the capacity reduction of link 5 is applied throughout its length, and not only at the downstream edge of the link, as with the macroscopic and mesoscopic models. This can be seen in Figure~\ref{fig:macromicro} as vehicles increase their spacing when they enter link 5. This causes congestion to form more quickly than in the previous case. The congestion wave reaches the model boundary after approximately 540 seconds.
Measuring from the plot, the speed of propagation of congestion in the microscopic region is approximately 8.4 km/hour, which is 8\% slower than in the fluid model, and faster than in the mesoscopic model. The queueing density in the microscopic model is about 34 vehicles in 500 meters. The congestion reaches the upstream boundary of the segment at around 1,400 sec. After that, the demand is turned off and congestion dissipates.

\begin{figure}[ht]
    \centering
    \includegraphics[width=0.9\textwidth]{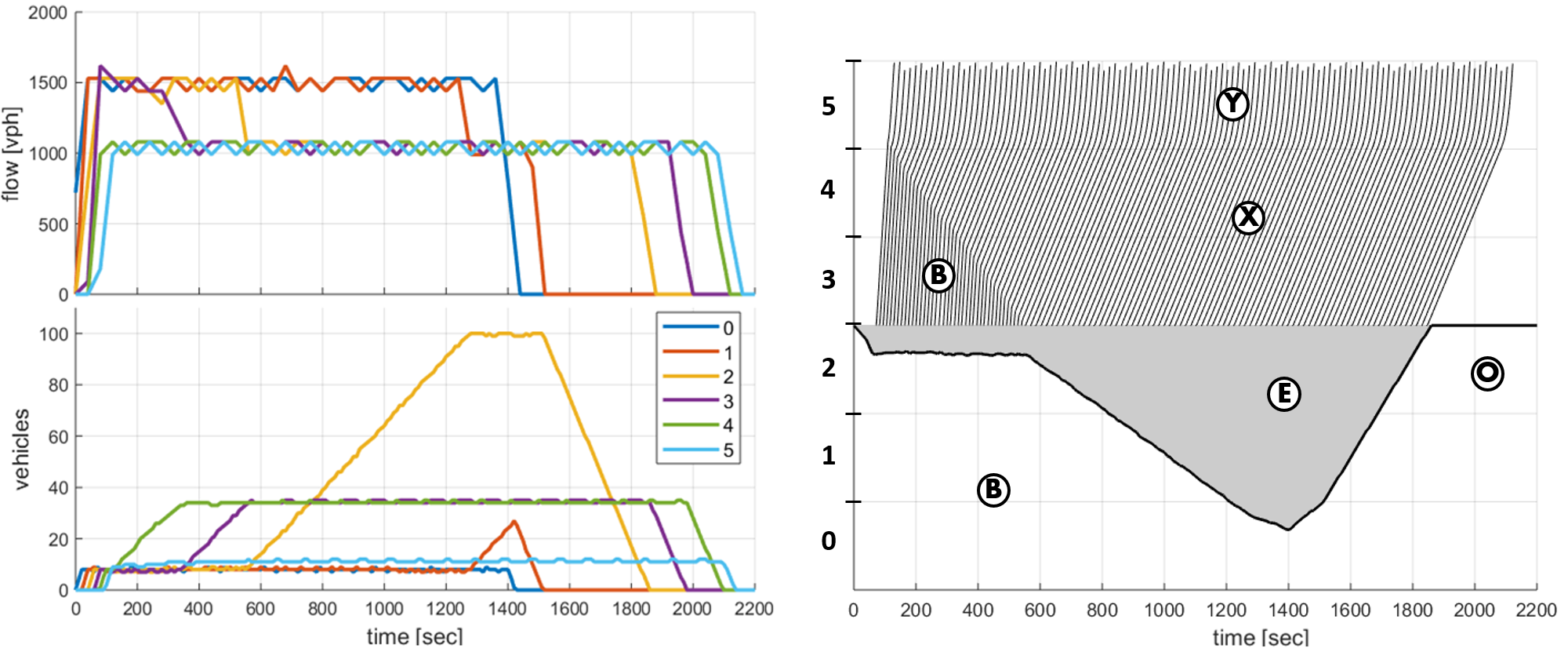}
    \caption{$m_A$=meso, $m_B$=micro.}
    \label{fig:mesomicro}
\end{figure}

\begin{figure}[ht]
    \centering
    \includegraphics[width=0.9\textwidth]{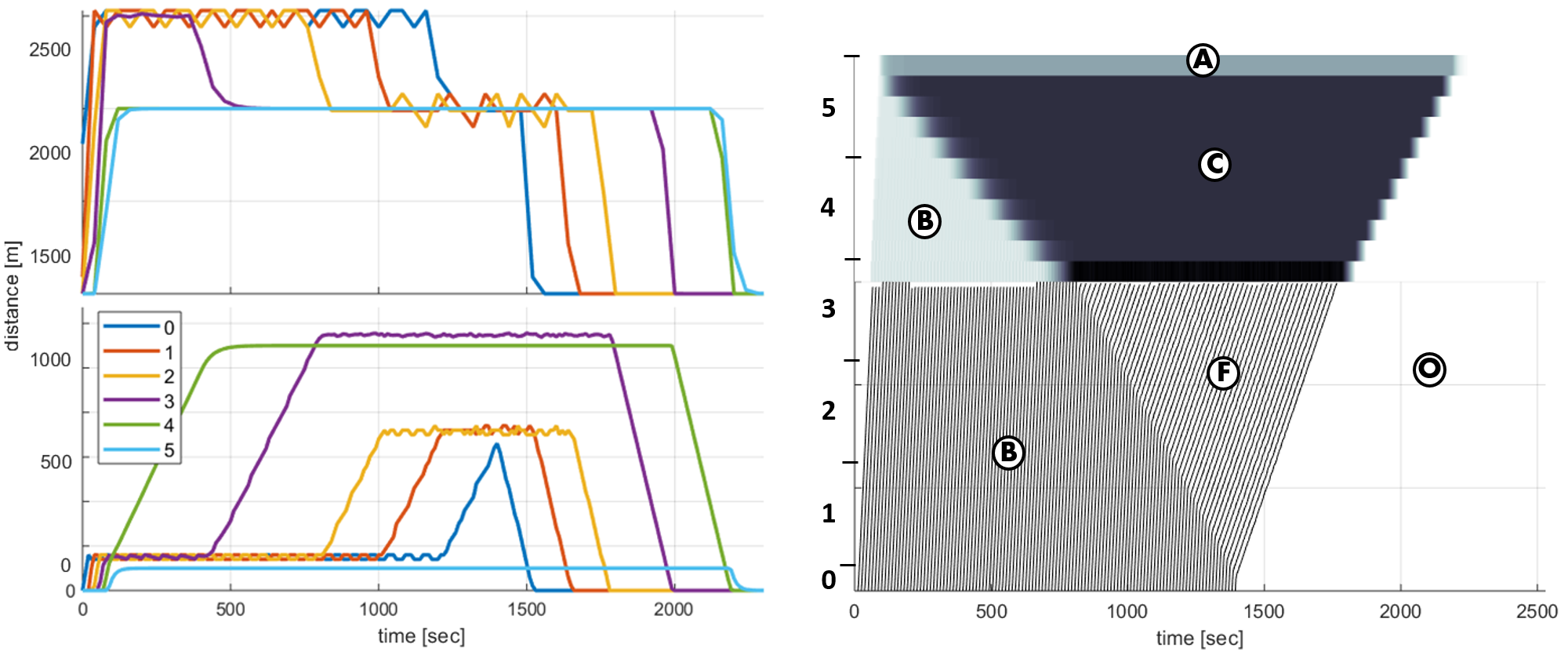}
    \caption{$m_A$=micro, $m_B$=macro.}
    \label{fig:micromacro}
\end{figure}

Figures \ref{fig:macromicro}, \ref{fig:mesomicro}, and \ref{fig:micromacro} show other combinations of the three model types. 

\section{Conclusion}
\label{sec:conclusions}
The goal of this work has been to develop a traffic simulation platform that is capable of running a variety of models, such as car-following, queueing, and fluid-based models. The approach has two main components. 
\begin{enumerate}
\item A network description that bridges the single-pipe graphs of macroscopic models and the lane-by-lane representations of micropscopic models. Road connections are used to construct lane groups. The lanes within a lane group are assumed to be synchronized in speed, and hence each lane group is assigned one instantiation of the `longitudinal dynamics'. This corresponds to a lane of vehicles for a microscopic model, a pair of queues for a mesoscopic model, and a differential equation for a macroscopic model. The granularity of the representation can be controlled by adding or removing road connections. Single-pipe and lane-by-lane are the two extreme cases.  
\item A protocol for coordinating models that operate on this network description. The protocol uses a `model interface' to negotiate the passage of flux packets from one link to another. The protocol is similar to the Godunov scheme for partial differential equations, with a simple extension to vehicle-based models.   
\end{enumerate}
Apart from the dynamical model, the complete specification of a traffic scenario also includes the demands and the control algorithms. The paper describes a demand specification that has multiple vehicles types (multi-commodity) and also allows for routed and probabilistic flows. The control structure has sensors and actuators, which provide communication between the models and the controllers. 

Open Traffic Models (OTM) is an implementation of this scheme. The software includes the three example models of the paper. The control and modeling interfaces allow for algorithms to be implemented as plugins. OTM is an open-source software, and it can be found at https://github.com/ggomes/otm-sim.


\end{document}